\documentclass{PoS}
\let\ifpdf\relax
\pdfoutput=1
\usepackage{amsmath}
\usepackage{amsfonts}
\usepackage{amssymb}
\usepackage{latexsym}
\usepackage{graphicx}
\usepackage{float}
\usepackage{wrapfig}

\newcommand{\be}{\begin{equation}}
\newcommand{\ee}{\end{equation}}

\newcommand{\tr}{{\rm tr\,}}
\newcommand{\Str}{{\rm Str\,}}

\newcommand{\Sdet}{{\rm Sdet\,}}

\newcommand{\UOSp}{{\rm UOSp\,}}

\newcommand{\IM}{{\rm Im\,}}
\newcommand{\eins}{\leavevmode\hbox{\small1\kern-4.8pt\normalsize1}}

\title{Random Matrix Models for the Hermitian Wilson-Dirac operator of QCD-like theories}

\ShortTitle{Random Matrix Models
 for $D_5$ of QCD-like theories}

\author{Mario Kieburg\\
        Department of Physics and Astronomy, State University of New York at Stony Brook, NY 11794-3800, USA\\
        E-mail: \email{mario.kieburg@stonybrook.edu}}

\author{Jacobus J. M. Verbaarschot\\
        Department of Physics and Astronomy, State University of New York at Stony Brook, NY 11794-3800, USA\\
        E-mail: \email{jv@chi.physics.sunysb.edu}}

\author{\speaker{Savvas Zafeiropoulos}\\
        Department of Physics and Astronomy, State University of New York at Stony Brook, NY 11794-3800, USA\\
        E-mail: \email{szafeiro@ic.sunysb.edu}}

\abstract{We introduce Random Matrix Models for the Hermitian Wilson-Dirac operator of QCD-like theories. We show that they are equivalent to the $\epsilon$-limit
of the chiral Lagrangian for Wilson chiral perturbation theory. Results are obtained for two-color QCD with quarks in the fundamental representation of the color group as well as any-color QCD with quarks 
in the adjoint representation. For $N_c=2$  we also have obtained the lattice spacing dependence of the quenched average spectral density for a fixed value of the index of 
the Dirac operator. Comparisons with direct numerical simulations of the random matrix ensemble are shown.}

\FullConference{The 30th International Symposium on Lattice Field Theory\\
                 June 24--29 ,  2012\\
                 Cairns, Australia}

\begin{document}

\section{Introduction}\label{sec1}
Chiral Random Matrix Theories \cite{SV,V} have been successful in describing lattice
QCD Dirac spectra on the scale of the eigenvalue spacing. It was shown
that they are equivalent to the $\epsilon$-limit of QCD which is given
by the $\epsilon$-limit of chiral perturbation theory \cite{OTV}. Recently, Random Matrix
Theory was extended to include discretization effects of both
the Wilson \cite{DSV} and the  staggered  
 Dirac operator \cite{James}. They
are equivalent to the $\epsilon$-limit of Wilson chiral perturbation theory
\cite{sharpe,rupak}
and staggered chiral perturbation theory, 
respectively \cite{bernard}. 

Starting from the chiral Lagrangian of Wilson chiral perturbation theory
in the microscopic domain, exact results were obtained for the 
spectral density of the Hermitian Dirac operator both for the quenched
case \cite{DSV} and the case of dynamical quarks \cite{SV11}. 
The spectral density of the
non-Hermitian Wilson Dirac operator, could only be accessed by means of
powerful random matrix techniques. The results for dynamical quarks show
that depending on the value of the low-energy constants either an
Aoki phase or a first order scenario is possible \cite{KSV}.
Lattice results for the eigenvalue density 
\cite{Damgaard:2011eg,Deuzeman:2011dh, heller}
have been compared successfully
to the exact results \cite{Akemann:2010zm,SV11,KVZ,KSV,ake-n}
for the spectral density in the microscopic limit.

Recently, a great deal of attention has been focused on the conformal
limit of QCD and QCD-like theories. 
 Both the two-color theory and the any color adjoint theory are 
relevant  for technicolor theories \cite{Rummukainen:2011xv}. 
The advantage of
$\rm {SU(2)}$ theories is that they  require less fermions
for achieving conformality  
than $\rm {SU(3)}$ theories.
 Furthermore, the $\rm {SU(2)}$ theory with two adjoint fermions  
is relevant  for minimal walking technicolor theories 
  \cite{francesco}. 
Studies of this theory have been performed for unimproved
Wilson fermions \cite{Bursa:2009we} and for the analysis of the conformal
window, it would be useful to have a better understanding of
the discretization errors.

As is the case in the continuum theory, also at non-zero lattice
spacing,  there is a one to one correspondence 
between patterns of chiral symmetry breaking and the 
anti-unitary symmetries of the Dirac operator \cite{V}.
We  therefore can distinguish three distinct classes. QCD in the fundamental
representation with three or more colors is the case
without anti-unitary symmetry. When we have an anti-unitary
symmetry,  $[T,D]=0$ for the Dirac operator $D$, then there are two different
possibilities. Either $T^2=1$ or $T^2 =-1$. In the first case it is always
possible to find a gauge field independent basis for which the Dirac
operator is real. This is the case for QCD with two colors in the fundamental
representation where the
chiral symmetry breaking pattern is  $\rm {SU(2N_{\rm f})}\rightarrow\rm 
{USp(2N_{\rm f})}$. In the second case it is possible to find a 
gauge field independent
basis in which the matrix elements are  expressed as self-dual quaternions.
This is the situation for QCD in the adjoint representation where 
the pattern of chiral symmetry breaking is $\rm {SU(2N_{\rm f})}\rightarrow\rm {SO(2N_{\rm f})}$.

The goal of this paper is to study the effect of a finite lattice spacing 
on the low lying Dirac eigenvalues and to understand the behavior 
of the spectral gap of $D_5+m\gamma_5=\gamma_5(D_{\rm W}+m)$ at finite 
quark mass as a function of the lattice spacing. In mean field theory, 
closure of the spectral gap will serve as an order parameter 
for the onset of the Aoki phase \cite{DSV}.

 This  paper is organized as follows. In section 2  we
 introduce a Wilson Random Matrix Model for  ${\rm SU}(2)$ with fundamental 
quarks and QCD with adjoint fermions. In particular, we consider 
the $N_{\rm f}$ flavor partition function as well as the 
partially quenched partition function. In section 3, we compare
analytical results with Monte Carlo data of the Random Matrix Model.

\section{Random Matrix Theory}

The random matrix for the Hermitian Wilson Dirac  operator proposed 
in  Refs.~\cite{DSV,Akemann:2010em} is given by 
\begin{equation}
D_5=\left(\begin{array}{cc}  aA &W\\W^{\dagger} & aB\end{array}\right),
\end{equation}
where $A$ and $B$ are Hermitian $ n\times n$ and $ (n+\nu) \times (n+\nu)$ matrices, and  $W$ is a complex $n\times (n+\nu)$  matrix. All matrix elements are 
distributed according to a Gaussian probability distribution.
The Random Matrix Theory for  $\beta=1$ and $\beta=4$  is obtained by
simply choosing real or quaternion matrix elements.

For  $a=0 $,  the Wilson-Dirac operator has  $ \nu$ generic zero modes  in accordance with the Atiyah-Singer index theorem. 
At finite  $a$,  
  one can define   the index of the 
Dirac operator for a fixed gauge field configuration
through spectral flow lines or equivalently by \cite{Itoh}
\begin{equation}
\label{defIndex}
\nu = \sum_{\lambda_k^W \in {\mathbb R}} {\rm sign} (\langle k
|\gamma_5|k\rangle),
\end{equation}
where the above sum is restricted to the real modes since the eigenfunctions corresponding to complex modes have zero chirality.

The  partition function of  $D_5$ with $ N_{\rm f}$ flavors  is given by
\begin{equation}
Z_{N_f}^{{\rm RMT},\nu}=\int dD_5 {\det}^{N_f} (D_5+m\gamma_5+z)P(D_5)\label{D5pf},
\end{equation}
where $P(D_5)$ is the probability distribution of the matrix elements of $D_5$.
In the microscopic limit where the combinations $\widehat{m}=2mn$ , $\widehat{z}=2zn$ and $\widehat{a}^2=a^2n/2$  are kept fixed as $n\to\infty$, the 
Random Matrix Theory reduces to the $\epsilon$-limit of Wilson chiral perturbation
theory,
\be
 Z^{\nu}_{ N_{\rm f}}=\int d\mu(U) det^{\kappa} U   
 \exp \left [\tr \frac{\widehat m}{2}  (U+U^{-1}) 
 -\tr \frac{\widehat z}{2}(U-U^{-1})+\widehat{a}^2\tr (U^2+U^{-2})\right]
\label{chiral}
 \ee 
 with $U\in {\rm U}(2N_f)/{\rm Sp}(2N_f)$ and $\kappa=\nu/2$ 
 for $\beta=1$ while $U \in {\rm U}(2N_f)/{\rm O}(2N_f)$ and $\kappa=\nu $ for $\beta=4$
as is the case for $a=0$ \cite{halasz-eff}.

The full partition function at fixed vacuum angle $\theta$ is 
given by a sum over the Fourier components $Z_{\nu}$,
\begin{equation}
Z(\theta) ~=~ \sum_{\nu=-\infty}^{\infty} ~e^{i\nu\theta} Z_{\nu} .
\end{equation}

In order to access the spectral properties of the Dirac operator 
we employ the supersymmetric method of RMT.
The generating function for the resolvent of $D_5$ is the partially quenched partition function
obtained by adding an additional 
fermionic and bosonic quark to the $N_f$ flavor partition function
  \begin{equation}
  \mathcal{Z}^\nu_{N_f+1|1}(\widehat{m},\widehat{z}, \widehat{z'};\widehat{ a}) =\displaystyle{ \left \langle {\det}^{N_f}(\gamma_5( D_{\rm W} + \widehat{m}))
\frac{\det(\gamma_5( D_{\rm W} + \widehat{m})+ \widehat{z}) }{\det(\gamma_5( D_{\rm W} + \widehat{m})+ \widehat{z'})}
\right \rangle_\nu.
}
\end{equation}
The resolvent of $D_5$ is given by
\be
 G^{\nu}(\widehat{z}, \widehat{m}; \widehat{a})=\displaystyle{\lim_{z'\to z}\partial_{z}
 \mathcal{Z}^{\nu}_{1|1}= \left< \tr\frac{1}{D_5+\widehat{z}} \right>}.
 \ee 
 Its discontinuity across the real axis gives the spectral density
\begin{equation}
 \rho_5^{\nu}(\widehat{\lambda}^5,\widehat{m};\widehat{a})=\frac{1}{\pi}\IM[G^{\nu}(\widehat{z}=\widehat{\lambda}^5,\widehat{m};\widehat{a})].
\end{equation}

In the microscopic limit the generating function reduces to a supersymmetric extension of
of the partition function (\ref{chiral}). However the integrals over the non-compact
part of $U$ are only convergent for imaginary $a$. To obtain an analytical continuation
to real $a$ we have to rotate  $U\rightarrow iU$. This results in the partition function
 \begin{eqnarray}
\mathcal{Z}^{\nu}_{ N_{\rm f}+1|1}&=&\int d\mu(U) \Sdet^{\kappa + N_{\rm f}}U \exp \left [\frac{i}{2}\Str \widehat{M} (U-U^{-1})\right] \nonumber  \\ 
 &\times&\exp \left [-\frac{i}{2}\Str \widehat{Z}(U+U^{-1})-\widehat{a}^2\Str (U^2+U^{-2})\right] .\label{pqpf}
 \end{eqnarray}
The  integration manifold $U\in\rm U \,( 2N_{\rm f}+2|2)/ \UOSp( 2N_{\rm f}+2|2)$  is the same as for  $a=0$. The mass matrix is given by
 $\widehat{M}= {\rm diag}(\widehat{m},\widehat{m}, \widehat{m}, \widehat{m})$ whereas the axial mass is
given by
 $\widehat{Z}={\rm diag}(\widehat{z}, \widehat{z}, \widehat{z'}, \widehat{z'})$.\\
 Note that for $z' \to z$ we recover the
   $ N_{\rm f}$ flavor partition function.

To evaluate this integral we need an explicit parameterization of $U$. Since it involves
only four Grassmann variables, it can easily be evaluated by brute force. The results are
quite lengthy and will be published elsewhere \cite{KVZ-2012}.
\begin{figure}[t!]
\begin{minipage}[b]{6.5cm}
\includegraphics[width=1.0\textwidth]{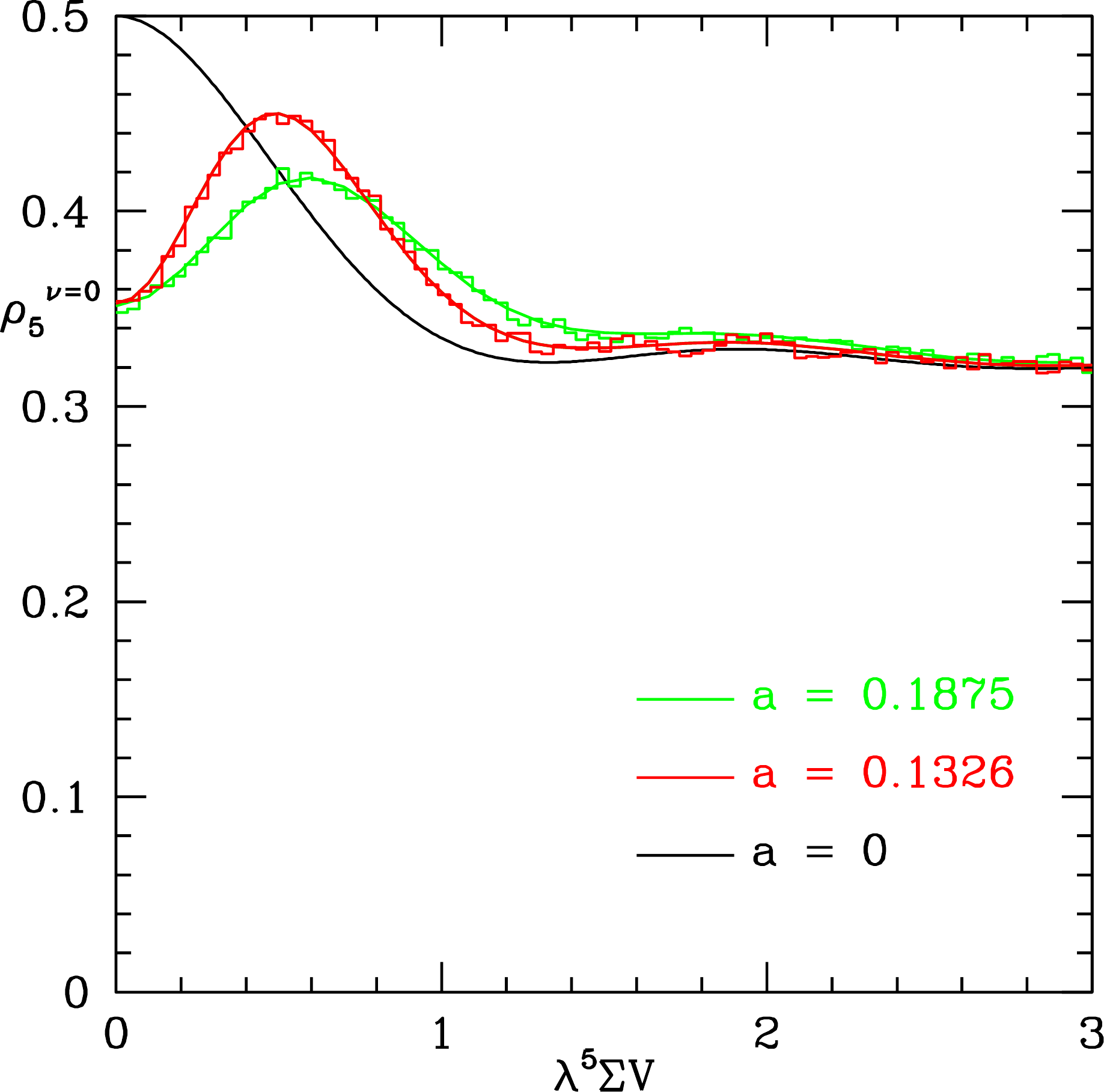}
\caption{The analytical results (solid curves) compared to the results of the Monte Carlo simulation (histograms).
}
\label{fig1}
\end{minipage}\quad
\begin{minipage}[b]{8.9cm}
\includegraphics[width=1.0\textwidth]{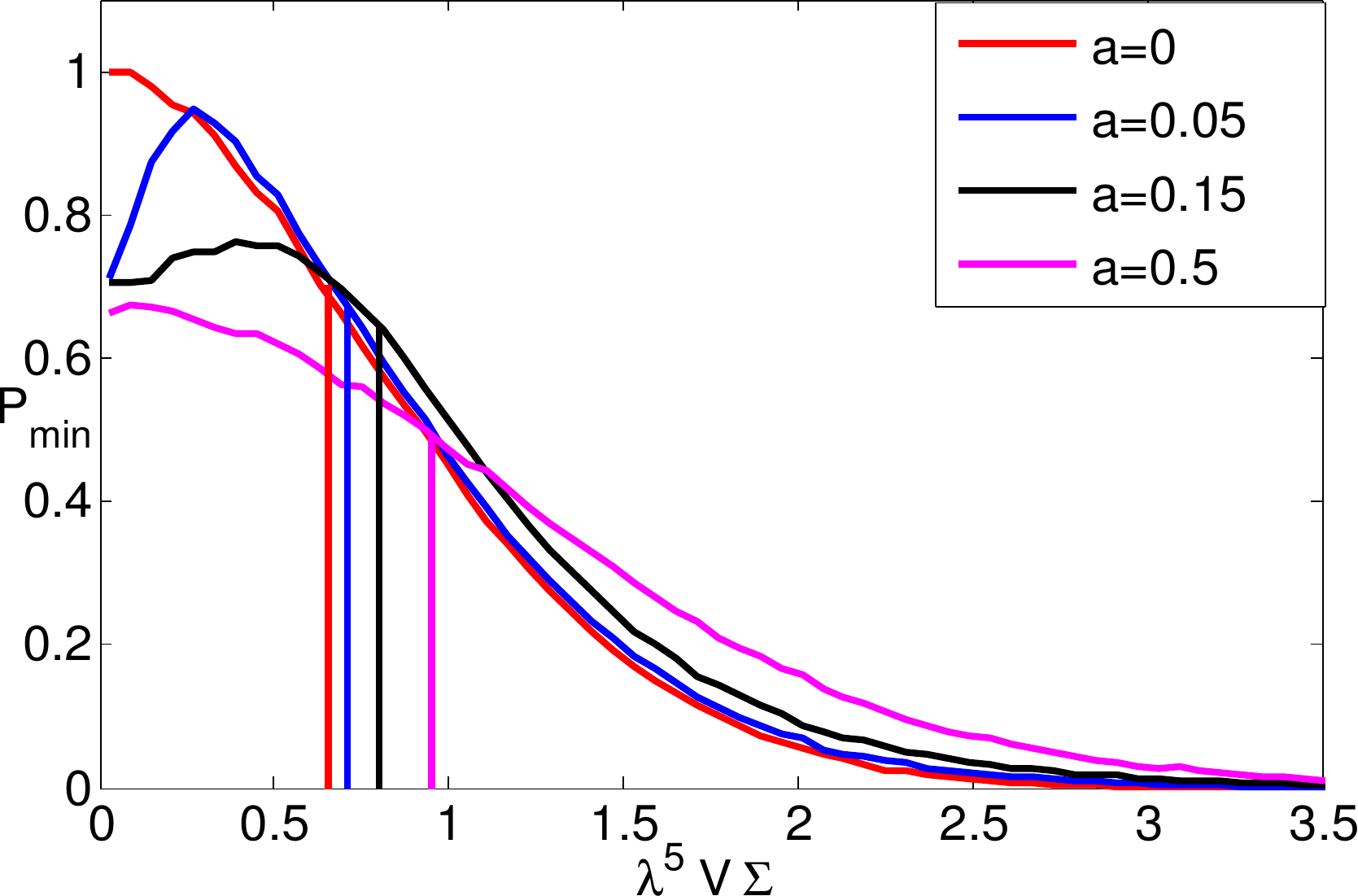}
\caption{ The distribution of the first positive eigenvalue,  with $\widehat{m}=0$ , $\nu=0$.
The average position of this eigenvalue (denoted by the vertical bar) 
shifts away from the origin for increasing $a$.  }
\label{fig2}
\end{minipage}
\end{figure}

\section{Analytical and Numerical results}\label{sec3}

To illustrate our analytical results we compare
in Fig.~\ref{fig1}  the results 
obtained from (\ref{pqpf}) for $\nu = 0$ and $m=0$ with  numerical
ones obtained by calculating the eigenvalues of an ensemble of random matrices. Surprisingly,
the spectral  
density at zero decreases by a factor $\sqrt 2$ for any nonzero value of $a$. 
 The reason for the nonuniformity of the $a \to 0$ limit is that for $a \ne 0$ the convergence
of the integral is achieved through the $U^2$-term while for $a =0$ the convergence comes from the $U$-term.
The effect of diagonal 
blocks that comprise the Wilson term in the Random Matrix Model can thus be seen
for arbitrarily small values of $a$.

In Fig.~\ref{fig2}  we study
the distribution of the first positive eigenvalue. Apparently, the diagonal blocks of
the Wilson Dirac operator lead to a weak repulsion of  the two eigenvalues 
closest to zero,  but for larger  
values of $a$, there is no repulsion away from zero. The average position of the first
eigenvalue increases as is shown by the corresponding vertical bar perpendicular to the
real axis.

\begin{figure}[t!]
\begin{minipage}[b]{7.2cm} 
\includegraphics[width=1.0\textwidth]{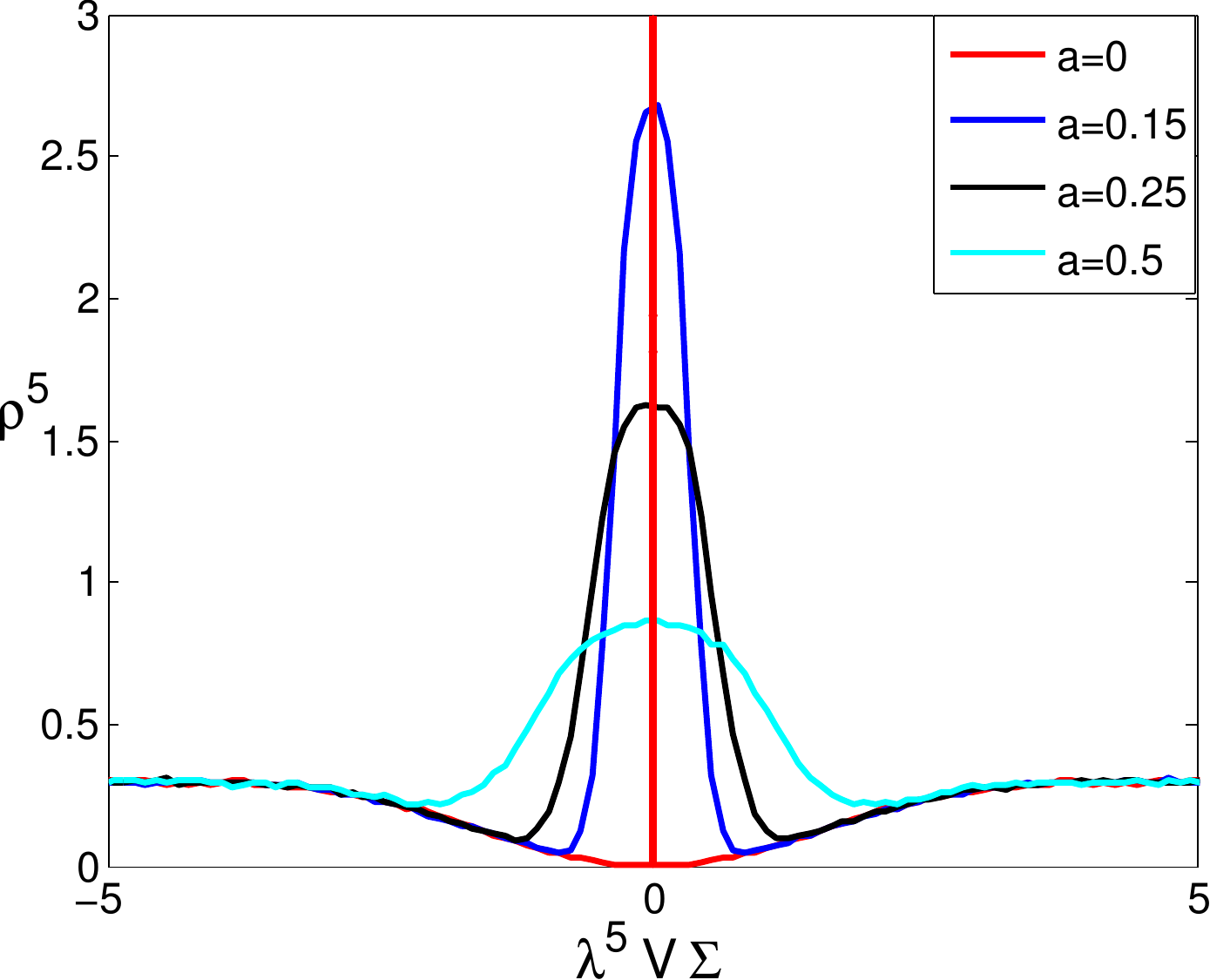}
\caption{Spectral density $\rho_5$ at $\nu=2$ and $\widehat{ m}=0$.
Note the presence of zero modes for 
$a=0$ and the widening of the peak as we increase $a$.}
\label{fig3}
\end{minipage}\quad
\begin{minipage}[b]{6cm} 
\includegraphics[width=1.0\textwidth]{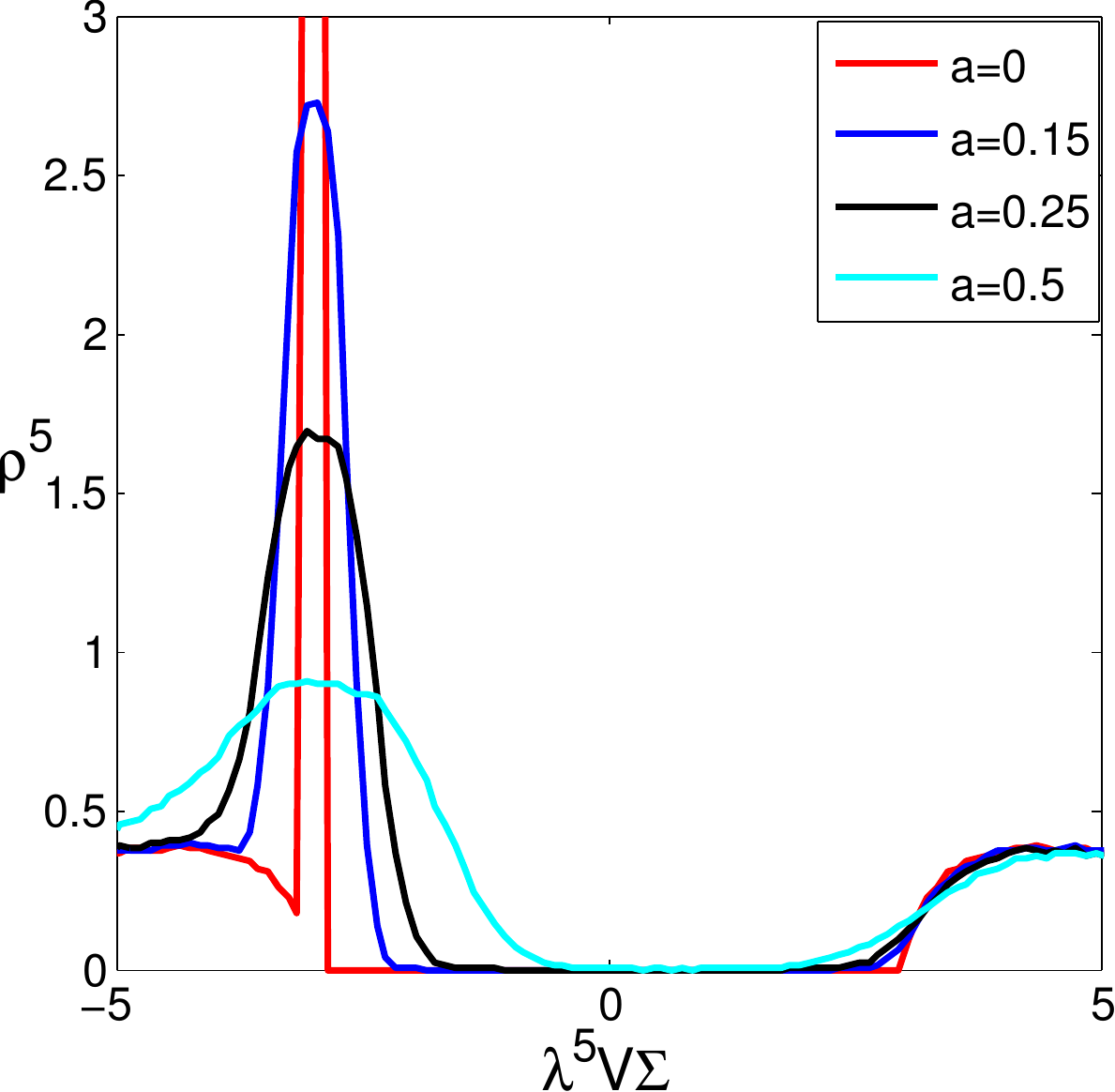}
\caption{Spectral density $\rho_5$ at $\nu=2$ and $\widehat{m}=3$.
For $a=0$ 
the spectrum has a  gap of width $2m$ which closes for increasing $a$.}
\label{fig4}
\end{minipage}
\end{figure}

For $\nu\neq 0$,         
the distribution of  
the zero modes is  
a Dirac delta function 
for $a=0$.  
As we increase the value of $a$ 
their distribution  
gets broadened with a width proportional to $a$.
This is similar to what happens in QCD with three or more colors in the fundamental representation. 
As can be seen from Figs.~\ref{fig3} and \ref{fig4}, at
about
 $\widehat{a}=0.5$ the peak due to the would be zero modes has disappeared almost completely.

For $a=0$ 
the spectrum of $D_5$ has a gap $[-m,m]$, but at finite lattice spacing 
eigenvalues of tail states penetrate 
the gap \cite{DSV}.  
 Our results provide an explicit analytical handle on these states
and allow us to identify the point where eigenvalues approach the center of the spectral gap
and inversion of the Dirac operator becomes very difficult.

For $ a= 0$, 
 the spectral density  of  the two-color theory  develops a square root 
type of singularity at the edge of the gap, 
$\rho_5(\widehat{\lambda}^5)\sim 1/\sqrt{(\lambda^5)^2-m^2}
+\nu\delta(\lambda^5-m)$
(see Fig. 4). 
For $\beta =2$, on the other hand, the spectral density approaches a finite limit at $\lambda^5=m$
. 

\begin{figure}[h!]
\begin{minipage}[b]{7.5cm}
\centering
\includegraphics[width=1.0\textwidth]{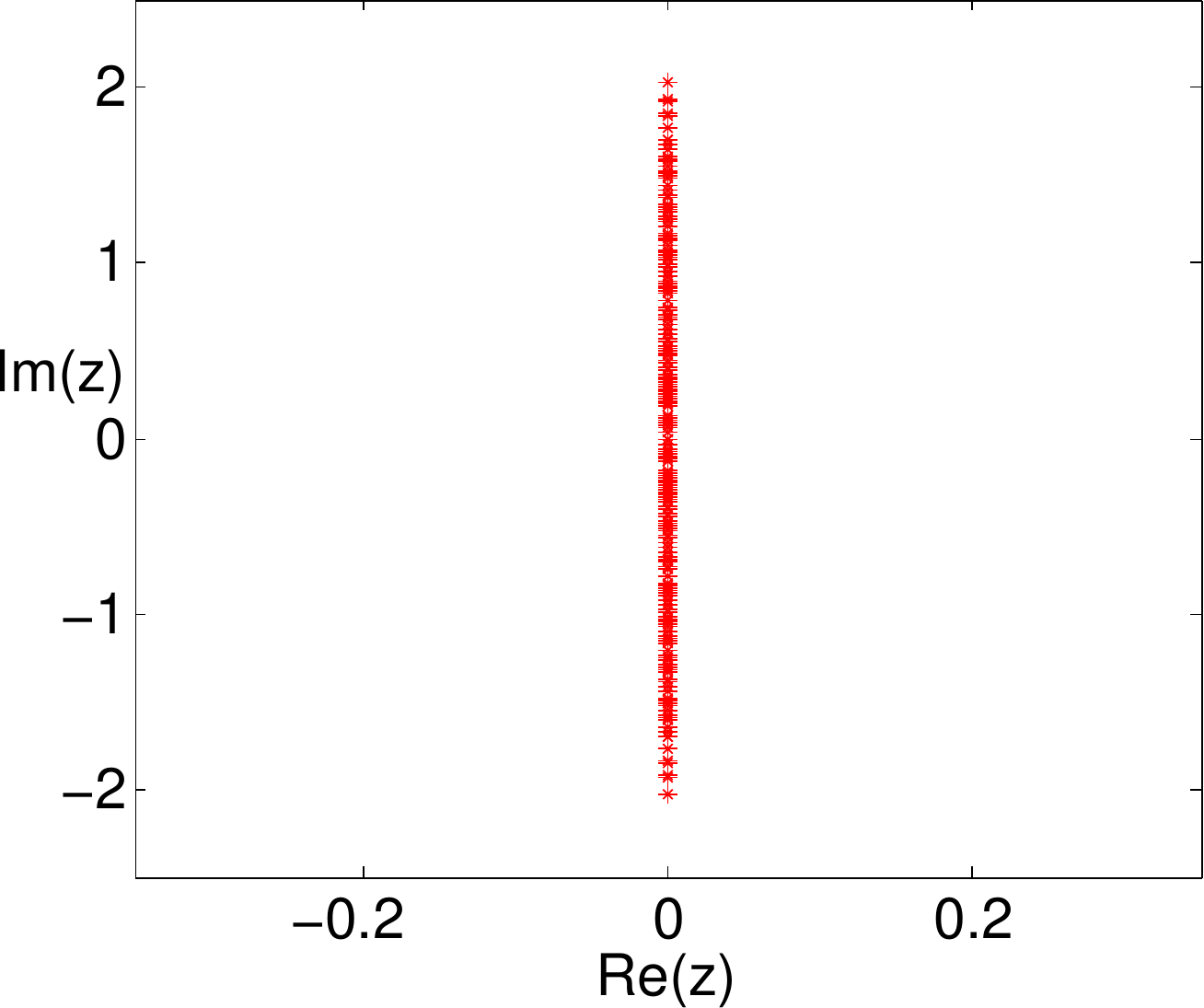}
\caption{Spectrum of a randomly generated matrix $D_{\rm W}$ with $\nu=5$ and $m=0$ at vanishing lattice spacing, i.e. $\widehat{a}=0$.}
\label{fig5}
\end{minipage}
\quad
\begin{minipage}[b]{7.2cm}
\centering
\includegraphics[width=1.0\textwidth]{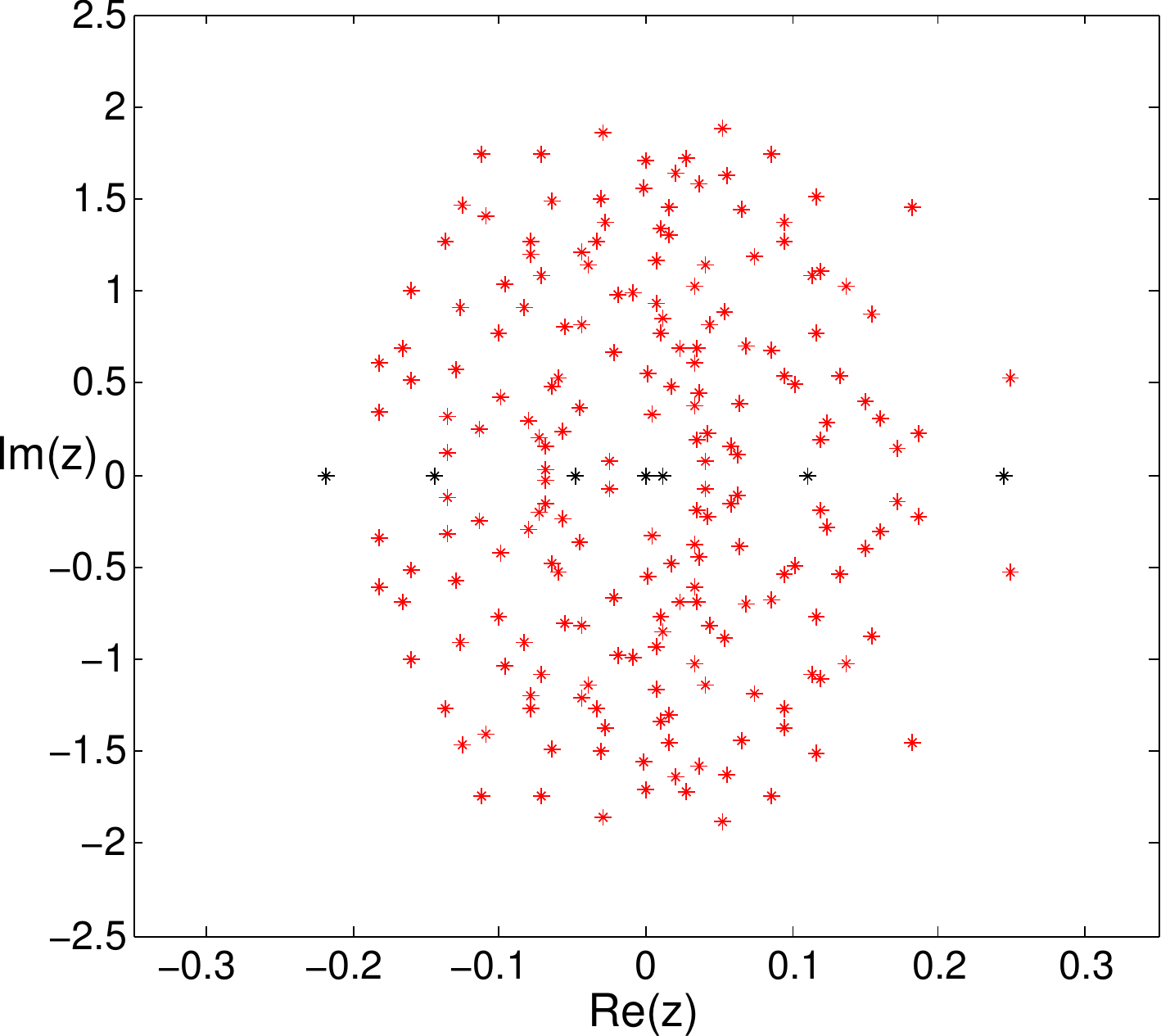}
\caption{Spectrum of a randomly generated matrix $D_{\rm W}$ with $\nu=5$ and $m=0$ for 
a finite lattice spacing ($\widehat{a}=1$).}
\label{fig6}
\end{minipage}
\end{figure}

In Figs.~\ref{fig5} and \ref{fig6} we show  scatter plots of the eigenvalues of $D_{\rm W}$. 
For $a=0$ 
the Wilson Dirac operator $D_{\rm W}$ 
is anti-Hermitian and the eigenvalues lie on the imaginary axis 
(see Fig.~\ref{fig5}). In contrast to this behavior, 
$D_{\rm W}$ is non-Hermitian at finite $a$. Hence, it has a complex spectrum. Because $D_{\rm W}$ is still
$\gamma_5$-Hermitian its complex eigenvalues occur 
in complex conjugate pairs, and at least $|\nu|$ eigenvalues are real (see Fig.~\ref{fig6}).
The additional real modes  always appear in an even number. In Fig.~\ref{fig6} 
we show a spectrum of $D_{\rm W}$ for $\nu = 5$ with  seven real modes.

\section{Conclusions}\label{sec4}

We have 
introduced Random Matrix Theories for the Wilson Dirac operator of QCD like theories 
and have 
obtained explicit analytical results for 
the spectral density of two-color QCD. 
The analytical results for $\nu=0$ have been 
compared to Monte Carlo simulations of the proposed random matrix ensemble. Furthermore, 
numerical results for the spectral density for non-zero quark mass and index 
$\nu$ have been 
presented. We stress that although the increase 
of computational power has allowed for lattice simulations in the deep 
chiral regime it is not possible to invert the Wilson Dirac operator when eigenvalues
are sufficiently close to zero. Our results identify the parameter domain where such eigenvalues
appear and can be potentially useful for identifying the parameter domain
for simulations with dynamical quarks.
Actually, the probability to obtain small eigenvalues is higher for 
the two-color theory 
than for QCD with more colors because of the lack of repulsion from the origin.
Our analytical results can be extended to the case 
of arbitrary $\nu$ and also to the case of adjoint QCD. 
These will be presented in  forthcoming publications.
 
\paragraph{Acknowledgments.}
 The authors are indebted to Kim Splittorff for his useful comments.
 MK acknowledges financial support by the Alexander-von-Humboldt Foundation. JV and SZ acknowledge support by
U.S. DOE Grant No. DE-FG-88ER40388.



\end{document}